\documentstyle[prb,aps,epsf]{revtex}
\begin{document}
\twocolumn[\hsize\textwidth\columnwidth\hsize
           \csname @twocolumnfalse\endcsname
\title{Quasiparticle transport equation with collision delay.
I. Phenomenological approach}

\author{V\'aclav \v Spi\v cka and Pavel Lipavsk\'y}
\address{Institute of Physics, Academy of Sciences, Cukrovarnick\'a 10,
16200 Praha 6, Czech Republic}
\author{Klaus Morawetz}
\address{Max-Planck Gesellschaft, Arbeitgruppe ``Theoretische
Vielteilchenphysik'' an der Universit\"at Rostock, D-18055 Rostock,
Germany}
\date{\today}
\maketitle
\begin{abstract}
For a system of non-interacting electrons scattered by resonant levels
of neutral impurities, we show that virial and quasiparticle
corrections have nearly equal magnitudes. We propose a modification of
the Boltzmann equation that includes quasiparticle and virial
corrections and discuss their interplay on a dielectric function.
\end{abstract}
\vskip2pc]
\section{Introduction}
Elastic scattering of electrons by impurities is the simplest but still
very interesting dissipative mechanism in semiconductors. Its simplicity
follows from the absence of the impurity dynamics, so that individual
collisions are described by a motion of an electron in a fixed
potential. On the other hand, due to a large variety of impurities
and their accessible concentrations, impurity-controlled transport
regimes span from simple response characterized by a mean-free path to
a weak localization.

Let us recall a quasiclassical picture of impurity controlled transport.
The basic effect of impurities on transport in crystals consists in
abrupt changes of directions of electron trajectories. Within the
Boltzmann equation (BE), this effect is described by scattering
integrals. At higher concentrations, impurities influence a band
structure. This correction can be built into the BE if one accepts that
elementary excitations are not simple electrons but electron-like
quasiparticles in the spirit of the Landau theory of Fermi liquids.
\cite{BP91} Finally, impurities attract/expel electrons to/from their
vicinity what reduces/increases a density of freely traveling electrons.
Such changes in effective density of electrons are covered by virial
corrections that are accounted for via non-local (in time and space)
corrections to scattering integrals. Although quasiparticle and virial
corrections to the BE are known for decades, transport theory that would
include both in the same time is still missing. Our aim is to fill this
gap. In this paper we focus on an intuitive approach. In the second
paper of this series we confirm equations presented here from the
quantum statistics.

To introduce basic concepts, we first discuss classical virial
corrections to the BE, Sec.~II. In particular, we show that non-locality
of scattering events can be described in terms of a collision delay.
In Sec.~III we review Wigner's concept of collision delay and estimate
magnitude of virial corrections for resonant levels in III-V
semiconductors. In Sec.~IV we show that virial corrections go
hand-in-hand with quasiparticle corrections having nearly equal
magnitudes. In Sec.~V, an intuitive modification of the BE and of the
most important observables (density, current, energy) is
proposed. In Sec.~VI, we discuss virial corrections to transport
coefficients. In Sec.~VII we summarize. In Appendix~A, we show that
non-self-consistent treatment used in this paper and more advanced
self-consistent treatment are equivalent within assumed precision.
In Appendix~B, we derive the derived optical theorem which explains
comparable magnitudes of the virial and the quasiparticle corrections.
In Appendix~C, we verify that the presented modification of the BE is
consistent with the equation of continuity and the energy conservation.

\section{Classical collision}
Since the quantum-mechanical theory with intuitively clear virial
corrections is still missing, the only experience for non-equilibrium
systems one can gain from virial corrections to the BE in the classical
statistical theory of moderately dense gases. Within accuracy to the
second order virial coefficient, these corrections were introduced
already on the break of centuries by Clausius.\cite{Chapman} Here we
modify his approach in two aspects. First, instead of binary collisions
of molecules we assume electron-impurity events. Second, instead of
space non-locality, we reformulate virial corrections in terms of time
non-locality.

According to Clausius,\cite{Chapman} one has to take into account that
two colliding molecules are not at the same space point, but at a
distance of sum of their radii. In other words, the scattering integral
has to be non-local in space.

Similar argument about non-locality of scattering event applies to
collisions of electrons with impurities. A sketch of a classical
trajectory of a colliding electron is in Fig.~\ref{fig:cl1}.
Before the electron reaches the impurity potential of a finite
range, its trajectory is a straight line. Then it makes a curve in the
impurity potential and again follows a straight line in a new direction.
Within the BE, this process is approximated by an effective event that
is local in time and space. Of course, within the local approximation,
one has to sacrifice dynamics of the electron during the collision. More
serious neglect follows from the fact that within the local
approximation the asymptotic motion along outgoing line cannot be
properly matched with the motion along the incoming line.

Let us find a correct matching of incoming and outgoing lines. To this
end, we extrapolate the incoming and outgoing lines and find their
crossover $X$. In general, such a crossover need not exist,
however, it always exists for spherical potentials to which we limit our
attention. The crossover $X$ gives us the coordinate at which we have to
place the effective event. As one can see in Fig.~\ref{fig:cl1}, the
crossover $X$ does not coincide with the centre of impurity. Using the
local scattering integral of the BE, the scattering event is placed in
the center of impurity, thus the shift of the center of scattering
event is the first neglect that influences motion of electron in the
asymptotic region.

The second neglect of the local approximation is not visible from the
trajectory in Fig.~\ref{fig:cl1}. The electron following the
extrapolated incoming line passes through the crossover $X$ at time
$t^{\rm in}$. The electron on the extrapolated outgoing line passes $X$
at time $t^{\rm out}$. In general, $t^{\rm in}\not=t^{\rm out}$,
however, the time locality of scattering integrals in the BE means that
$t^{\rm in}=t^{\rm out}$.

One finds two misfits: in the position of the event, and in the
matching of times. Although these two misfits usually come together,
their consequences can be discussed separately by using special models.

\subsection{Point traps}
First, we assume impurities of a negligible volume with a capability to
trap electrons for a certain time. In this case, there is no misfit in
the position but $t^{\rm out}-t^{\rm in}=\Delta_t>0$.

Using intuitive arguments, the collision delay $\Delta_t$ can be
incorporated into the scattering integrals of the BE. A balance equation
of the Boltzmann type for scattering by impurities
reads
\begin{eqnarray}
{\partial f\over\partial t}+{k\over m}{\partial f\over\partial r}-
{\partial\phi\over\partial r}{\partial f\over\partial k}
&=&
\int{dp\over(2\pi)^3}P_{pk}f\left(p,r,t^{\rm in}_{pk}\right)
\nonumber\\
&-&
\int{dp\over(2\pi)^3}P_{kp}f\left(k,r,t^{\rm in}_{kp}\right),
\label{cl1}
\end{eqnarray}
where $f(k,r,t)$ is a distribution function in the phase space, $r$ is a
coordinate, $t$ is a time, $k$ and $p$ are momenta, $P_{kp}$ is the
scattering rate from $k$ to $p$. Since distributions in the scattering
integrals correspond to initial conditions, $t^{\rm in}_{kp}$
is a time at which an electron enters the scattering from $k$ to $p$.

In the scattering-out event [the second term on the r.h.s. of
(\ref{cl1})], an electron of momentum $k$ enters a collision at
$t^{\rm in}_{kp}$ leaving at $t^{\rm out}_{kp}$ with momentum $p$. The
scattering-out integral gives a probability that at time $t$ an electron
leaves the momentum $k$. This happens at the beginning of the collision,
thus $t^{\rm in}_{kp}=t$.

In the scattering-in event [the first term on the r.h.s of
(\ref{cl1})], an electron of momentum $p$ enters a collision at
$t^{\rm in}_{pk}$ leaving at $t^{\rm out}_{pk}$ with momentum $k$.
The scattering-in integral gives a probability that at time $t$ an
electron enters the momentum $k$. This happens at the end of the
collision, thus $t^{\rm out}_{pk}=t$. From
$t^{\rm out}_{pk}-t^{\rm in}_{pk}=\Delta_t(p,k)$, one finds that
$t^{\rm in}_{pk}=t-\Delta_t(p,k)$.
The time argument in the scattering-in is thus shifted by the collision
delay $\Delta_t(p,k)$. A modified BE then reads
\begin{eqnarray}
{\partial f\over\partial t}+{k\over m}{\partial f\over\partial r}-
{\partial\phi\over\partial r}{\partial f\over\partial k}
&=&
\int{dp\over(2\pi)^3}P_{pk}f(p,r,t-\Delta_t)
\nonumber\\
&-&
\int{dp\over(2\pi)^3}P_{kp}f(k,r,t).
\label{cl2}
\end{eqnarray}

Electrons trapped by impurities are excluded from free motion. With
a finite collision delay, one has to deal with two distinguished local
densities of electrons. Beside the physical density $n=N/\Omega$ (number
of electron $N$ per volume $\Omega$), there is an effective density
\begin{equation}
n_{\rm free}(r,t)=\int{dk\over(2\pi)^3}f(k,r,t)
\label{cl3}
\end{equation}
which equals the local density in the free space between impurities.

For finite collision delay $\Delta_t$, a share of electrons trapped by
impurities can change in time. Accordingly, the free density $n_{\rm free}$
does not conserve. From (\ref{cl2}) one finds that in a homogeneous but
non-stationary system
\begin{eqnarray}
{\partial n_{\rm free}\over\partial t}
&=&
\int{dk\over(2\pi)^3}{dp\over(2\pi)^3}P_{pk}
\left(f(p,t-\Delta_t)-f(p,t)\right)
\nonumber\\
&=&-\int{dk\over(2\pi)^3}{dp\over(2\pi)^3}P_{pk}
\Delta_t{\partial f(p,t)\over\partial t}
\nonumber\\
&=&-{\partial\over\partial t}
\int{dk\over(2\pi)^3}{dp\over(2\pi)^3}P_{pk}\Delta_t f(p,t).
\label{cl4}
\end{eqnarray}
The quantity that conserves is the physical density
\begin{equation}
n=n_{\rm free}+n_{\rm corr}
\label{cl5a}
\end{equation}
which differs from the free density by the density
\begin{equation}
n_{\rm corr}=\int{dk\over(2\pi)^3}{dp\over(2\pi)^3}P_{pk}\Delta_t f(p,t)
\label{cl5}
\end{equation}
that is correlated with impurity positions.

Note that the scattering mechanism enters relation between density $n$
and distribution $f$. Without virial correction (here represented by
correlated density), the functional $n[f]$ is independent of scattering,
since $n=n_{\rm free}$. In the presence of virial corrections one has to
keep in mind that a density of freely traveling electrons does not equal
the physical density.

\subsection{Hard spheres}
As the second example, we discuss hard-sphere impurities. In this case,
the incoming and outgoing lines have a crossover at the sphere surface.
The times match exactly, $t^{\rm out}=t^{\rm in}$. The only misfit
results from the fact that the crossover is not at the centre of
the impurity but shifted by the sphere radius. Here we show that the
crossover offset can be reformulated in terms of an effective time
mismatch so that one can use unified description of collisions with
point traps and with hard spheres.

A collision with a hard sphere is schematically shown in
Fig.~\ref{fig:cl2}. The real electron trajectory follows the full
line. The scattering integral of the BE describes this event by an
electron following the dashed line. This effective trajectory (from
$\bar{\rm A}$ to $\bar{\rm B}$) is longer than the real one (from A to
B) by $\Delta_s=2\sqrt{R^2-b^2}$.

One can include the finite size of impurities into the transport
equation in a manner to parallel traps. We approximate the trajectory
of the electron by the effective trajectory $\bar A\bar B$. Since,
following the a real trajectory AB, the electron reaches a next
collision sooner by a time $\Delta_s/u$, we introduce into transport
equation (\ref{cl2}) a negative time delay $\Delta_t=-\Delta_s/u$.
Here $u$ is an electron velocity.

For the hard-sphere impurities, transport equation (\ref{cl2}) with
the negative collision delay is only an approximation. Let us check
how this approximation works for the correlated density. The classical
scattering rate on hard spheres reads
\begin{equation}
P_{pk}={(2\pi)^3\over k^2}\delta(|k|-|p|)
c^\prime u{R^2\over 4}\sin\vartheta,
\label{clh1}
\end{equation}
where $c^\prime=N_{\rm imp}/\Omega$ is an impurity concentration (number
of impurities $N_{\rm imp}$ per volume $\Omega$), $u=k/m$ is an electron
velocity, and $\vartheta$ is a scattering angle,
$pk=|k||p|\cos\vartheta$. The inverse lifetime follows from (\ref{clh1})
as
\begin{equation}
{1\over\tau}=\int{dp\over(2\pi)^3}P_{pk}=c^\prime u\pi R^2.
\label{clh2}
\end{equation}

The collision delay $\Delta_t=-\Delta_s/u$ in angular coordinates reads
\begin{equation}
\Delta_t=-{2\over u}\sqrt{R^2-b^2}=-{2\over u}R\sin{\vartheta\over 2}.
\label{clh3}
\end{equation}
The correlated density from (\ref{cl5}) results
\begin{eqnarray}
n_{\rm corr}
&=&-\int{dp\over(2\pi)^3}f(p,r,t)
\nonumber\\
&&\ \ \ \times
{1\over(2\pi)^3}\int_0^\infty k^2dk\int_0^{2\pi}d\varphi
\int_0^\pi d\vartheta P_{pk}\Delta_t
\nonumber\\
&=&-c^\prime {4\pi\over 3}R^3\int{dp\over(2\pi)^3}f(p,r,t)
\nonumber\\
&=&-c^\prime {4\pi\over 3}R^3 n_{\rm free}
\nonumber\\
&=&-{\Omega_{\rm imp}\over\Omega} n_{\rm free}.
\label{clh4}
\end{eqnarray}
Here, we have denoted
$\Omega_{\rm imp}=\Omega c^\prime{4\pi\over 3}R^3=N_{\rm imp}
{4\pi\over 3}R^3$ the total volume of impurities.

The physical content of the correlated density can be demonstrated on
the equation of state. The number of electrons which hit the surface
of the sample is given by the density of freely traveling electrons
$n_{\rm free}$, therefore the pressure $P$ is given by the equation of
state
\begin{equation}
P=n_{\rm free}k_BT.
\label{clh5a}
\end{equation}
>From (\ref{cl5a}) and (\ref{clh4}) we find relation between the free
density and the total number of particles $N$
\begin{equation}
n_{\rm free}={N\over\Omega-\Omega_{\rm imp}}.
\label{clh6}
\end{equation}
The equation of state (\ref{clh5a}) thus takes the form of the van der
Waals equation
\begin{equation}
P(\Omega-\Omega_{\rm imp})=Nk_BT.
\label{clh5}
\end{equation}
Briefly, the negative collision delay simulates for the excluded volume
in the van der Waals equation of state.

Note that the correlated density (\ref{clh4}) is negative. The
density $n_{\rm free}$ in the free space between impurities is higher
than the physical density $n$ what reflects that electrons are
expelled from the volume of impurities. It is important to distinguish
which density ($n_{\rm free}$ or $n$) is relevant for individual
physical quantities. For instance, the charge density is given by $n$,
while pressure relates to $n_{\rm free}$.

\section{Collision delay time in quantum mechanics}
The classical statistics shows that the non-locality of scattering
events is approximatively described by the collision delay. This
concept is easily transferred to the quantum mechanics, where the
collision delay has already been introduced by Wigner.\cite{W55} He
used the maximum of wave packet to identify motion of an electron.
Now we apply Wigner's approach to a neutral impurity to estimate a
magnitude of virial corrections.

The scattering of electron by a single impurity is described by the
Schr\"odinger equation\cite{M61}
\begin{equation}
(\omega-H_0-V)(\psi_{\rm in}+\psi_{\rm out})=0,
\label{ts1}
\end{equation}
where $\psi_{\rm in}(r)=\exp{ikx}$ is an incoming plane wave,
with $r\equiv(x,y,z)$,
$\psi_{\rm out}$ is the outgoing part, $H_0$ is Hamiltonian of unperturbed
crystal and $V$ is the impurity potential. The incoming plane wave has
to be an eigen state of the crystal, $(\omega-H_0)\psi_{\rm in}=0$, thus
the energy equals the kinetic energy of the incoming plane wave,
$\omega=\epsilon_k$. Then ({\ref{ts1}) simplifies as
\begin{equation}
(\epsilon_k-H_0)\psi_{\rm out}=V(\psi_{\rm in}+\psi_{\rm out}).
\label{ts2}
\end{equation}

A formal solution of equation (\ref{ts2}) reads\cite{M61}
\begin{equation}
\psi_{\rm out}=G_0^R(\epsilon_k)T^R(\epsilon_k)\psi_{\rm in},
\label{ts3}
\end{equation}
where
\begin{equation}
G^R_0(\omega)={1\over \omega-H_0+i0},
\label{ts4}
\end{equation}
is the retarded Green's function of the host crystal, and
\begin{equation}
T^R=V+VG_0^RT^R,
\label{ts5}
\end{equation}
is the T-matrix.

As a model potential of the neutral impurity we use the one proposed
by Koster and Slater\cite{KS54a,KS54b}
\begin{equation}
V=|0\rangle v\langle 0|,
\label{ts6}
\end{equation}
where $|0\rangle$ is a single orbital at the impurity site. We will use
the convention that lowercase denotes local elements of operators (that
are in uppercase) throughout the paper. For the Koster-Slater
potential, the T-matrix is also restricted to the selected orbital,
$T^R=|0\rangle t^R\langle 0|$, and reads
\begin{equation}
t^R=v+v\langle 0|G_0^R|0\rangle t^R=
{v\over 1-v\langle 0|G_0^R|0\rangle}.
\label{ts7}
\end{equation}

To obtain the collision delay, we place the impurity in the initial of
coordinates and express the wave function  in the time
representation
\begin{equation}
\psi(r,t)={\rm e}^{ikx-i\epsilon_kt}-{m\over 2\pi|r|}t^R(\epsilon_k)
{\rm e}^{ik|r|-i\epsilon_kt}.
\label{ct1}
\end{equation}
We have used an asymptotic Green's function for
large $r$, see Ref.~\onlinecite{M61},
\begin{equation}
\langle r|G^R_0(\epsilon_k)|0\rangle =-{m\over 2\pi|r|}{\rm e}^{ik|r|},
\label{ts9}
\end{equation}
to evaluate the outgoing wave from (\ref{ts3}). This approximation holds
for energies $\epsilon_k$ in the parabolic region of the band structure,
$\epsilon_k=k^2/2m$. The first term in (\ref{ct1}) is the incoming wave
$\psi_{\rm in}$ and the second one is the outgoing part
$\psi_{\rm out}$.

To see the time delay, we take a linear combination of wave functions
$\psi$ so that the incoming part $\psi_{\rm in}$ forms a wave packet of
a narrow momentum width $\kappa\to 0$,
\begin{eqnarray}
\psi_{\rm in}(r,t)
&=&{1\over\sqrt{\pi}\kappa}
\int dp {\rm e}^{-{(p-k)^2\over\kappa^2}}
{\rm e}^{ipx-i\epsilon_pt}
\nonumber\\
&\approx &{\rm e}^{ikx-i\epsilon_kt}
\exp\left\{-{\kappa^2\over 4}(x-ut)^2\right\},
\label{ct3}
\end{eqnarray}
where $u=k/m$ is an electron velocity.
This wave packet passes the initial of coordinates at $t=0$. A
corresponding outgoing wave $\psi_{\rm out}$ reads
\begin{eqnarray}
\psi_{\rm out}(r,t)&=&-{m\over 2\pi|r|}
{1\over\sqrt{\pi}\kappa}
\int dp {\rm e}^{-{(p-k)^2\over\kappa^2}}
t^R(\epsilon_p){\rm e}^{ip|r|-i\epsilon_pt}
\nonumber\\
&\approx &-{m\over 2\pi|r|}t^R(\epsilon_k)
{\rm e}^{ik|r|-i\epsilon_kt}
\nonumber\\
&\times &
\exp\left\{-{\kappa^2\over 4}\left(|r|-u\left(t+
\left.{i\over t^R}{\partial t^R\over\partial\omega}
\right|_{\omega=\epsilon_k}\right)\right)^2\right\}.
\nonumber\\
\label{ct4}
\end{eqnarray}
The outgoing wave passes the initial of coordinates with the collision
delay
\begin{equation}
\Delta_t={\rm Im}\left.{1\over t^R}{\partial t^R\over\partial\omega}
\right|_{\omega=\epsilon_k}.
\label{ct5}
\end{equation}
The collision delay (\ref{ct5}) depends only on the energy of electron.
This is because the Koster-Slater impurity has a single scattering
channel of the s-symmetry. For a general potential $V$, collision delay
$\Delta_t$ depends also on the scattering angle as the classical
collision delay (\ref{clh3}).

The collision delay (\ref{ct5}) is a quantum counterpart of the
classical collision delay (\ref{clh3}). Following analogy between
the quantum and classical approaches to the Boltzmann-like transport
equations, we introduce the collision delay (\ref{ct5}) into the
scattering integral in exactly the same way as in the classical
case. In other words, we expect the transport equation to be of form
(\ref{cl2}), however, scattering rates $P_{kp}$ and collision
delay $\Delta_t$ are extracted from quantum collisions.

The rate of scattering by Koster-Slater impurities of concentration $c$
(probability that impurity occupy a site) follows from the Fermi golden
rule as
\begin{equation}
P_{pk}=c|t^R(\epsilon_k)|^2 2\pi\delta(\epsilon_k-\epsilon_p).
\label{ct6a}
\end{equation}
This scattering rate does not depend on the scattering angle, thus
it can be also expressed in terms of the lifetime $\tau$
\begin{equation}
P_{pk}={1\over\tau}{2\pi^2\over k^2}\delta(|p|-|k|),
\label{ct6}
\end{equation}
where $\tau$ is conveniently evaluated from the T-matrix
\begin{equation}
{1\over\tau}=c(-2){\rm Im}t^R(\epsilon_k).
\label{ct7}
\end{equation}

\subsection{Estimate of virial corrections}
>From a scattering by a single impurity one can estimate magnitude of
virial corrections. Using formula (\ref{cl5}) with the quantum
scattering rate (\ref{ct6}) and collision delay (\ref{ct5}), one finds
correlated density
\begin{equation}
n_{\rm corr}=\int {dk\over(2\pi)^3}f(k){\Delta_t\over\tau}.
\label{ct7a}
\end{equation}
The magnitude of virial corrections is thus measured by a ratio
$\Delta_t/\tau$.

Note that the collision delay is independent from the impurity
concentration, while the lifetime is inversely proportional to the
concentration. Accordingly, $\Delta_t/\tau\sim c$, i.e., magnitude of
virial corrections is controlled by the impurity concentration $c$. To
be specific, we will assume impurity concentrations $\sim 10^{-6}$ per
site.

Now we estimate $\Delta_t/\tau$ for a model local Green's function
\cite{SLV92}
\begin{eqnarray}
&&\langle 0|G_0^R(\omega)|0\rangle
\nonumber\\
&=&{2\over W}\left(-{b_1\over 2}-{b_3\over 8}+z+\left(b_1-
{b_3\over 2}\right)z^2+b_3z^4\right)
\nonumber\\
&+&\left.\theta(1-z^2){2\over W}
(1+b_1z+b_3z^3)\sqrt{1-z^2}\right|_{z={\omega\over W}-1}.
\label{ts8}
\end{eqnarray}
Here, $W=6$~eV is a half-width of a conductivity band, and parameters
$b_1=1.2$ and $b_3=-.4$ serve to model the local density of state to
a shape resembling III-V semiconductors, see Fig.~\ref{fig:ts1}.

The collision delay is very sensitive to a value of the impurity
potential $v$. Using (\ref{ts7}), one can rearrange the collision
delay (\ref{ct5}) as
\begin{equation}
\Delta_t=-{\rm Im}\left[t^R{\partial\over\partial\omega}{1\over t^R}
\right]={\rm Im}{v{\partial\over\partial\omega}
\langle 0|G^R_0|0\rangle\over1-v\langle 0|G^R_0|0\rangle}.
\label{ts7a}
\end{equation}
Apparently, the collision delay will be long for potentials for which
the denominator $1-v\langle 0|G_0^R|0\rangle$ goes to zero. For these
values of potential $v$, the impurity behaves like a resonant level
close to the conductivity band edge.

For model function (\ref{ts8}), the real part of the local Green's
function at the band edge $\omega=0$ equals to $-0.185$~1/eV. For
potentials $v<-5.4$~eV, an impurity has a bound state. For $v>-5.4$~eV,
there is a resonant level. In our calculations we use value
$v=-5.35$~eV. In Fig~\ref{fig:ct1}, the energy dependence of $\Delta_t$
evaluated from (\ref{ts7a}) is compared with the lifetime $\tau$ from
(\ref{ct7}). In Fig~\ref{fig:ct1} one can see that
$\Delta_t/\tau\sim 0.1$, therefore appreciable virial corrections appear
already for assumed concentration of resonant level $c=10^{-6}$ per
site.

The strong dependence of the collisional delay on the position of
the resonant level leads to a strong dependence of virial corrections
on the impurity potential, see Fig.~\ref{fig:vq2}. Such changes of the
impurity potential can be achieved for instance by a hydrostatic
pressure.\cite{Kristofik} The impurity concentration and the
hydrostatic pressure can be thus used to control magnitude of virial
corrections.

\section{Quasiparticle picture}
>From analysis of the scattering by the Koster-Slater impurity, we have
found that the largest virial corrections appear for resonant levels.
Resonant levels, however, also result in large values of the T-matrix,
as one can see in Fig.~\ref{fig:ts2}. At the band edge
${\rm Re}t^R\sim -400$~eV and ${\rm Im}t^R\sim-30$~eV. In particular,
the real part of the T-matrix is large compared to potential
$v=-5.35$~eV. For such large values of the T-matrix, the impurity
scattering affects the electronic band structure. To take this effect
into account we have to treat electrons as quasiparticles.

\subsection{Averaged T-matrix approximation}
The multiple scattering by impurities has been described in detail
already in 1960's within Green's functions.\cite{VKE68} In the averaged
T-matrix approximation (ATA) that corresponds to our approximation of
scattering rates, the self-energy equals the averaged value of the
T-matrix,
\begin{equation}
\Sigma^R=cT^R.
\label{qp5}
\end{equation}
Since $T^R=\sum_r|r\rangle t^R(r)\langle r|$, we can write the
self-energy as
\begin{equation}
\Sigma^R=\sum_r|r\rangle \sigma^R(r)\langle r|.
\label{qp5a}
\end{equation}

\subsection{Energy renormalization}
The quasiparticle energy that describes propagation given by effective
``Hamiltonian'' $H_0+cT$ reads
\begin{equation}
\varepsilon_k=\epsilon_k+{\rm Re}\sigma^R(\epsilon_k).
\label{qp7}
\end{equation}
The imaginary part of the self-energy provides the life time
\begin{equation}
{1\over\tau}=-2{\rm Im}\sigma^R(\epsilon_k),
\label{qp5b}
\end{equation}
which is identical to the Fermi golden rule value (\ref{ct6}).

In Fig.~\ref{fig:qp1} one can see that effect of impurities with the
resonant level on the band structure is rather profound. The major
effect is the overall shift of the band. This shift does not influence
bulk properties of homogeneous crystals because it is compensated by
a shift of chemical potential.

The energy renormalization leads to quasiparticle corrections to
velocity
\begin{equation}
u={\partial\varepsilon_k\over\partial k}\not= {k\over m}.
\label{qp9}
\end{equation}
Taking the momentum derivative from (\ref{qp7}) one finds the
renormalized velocity as
\begin{equation}
u=z{k\over m},
\label{qp10}
\end{equation}
where $z$ is the wave-function renormalization
\begin{equation}
z(k)=1+\left.{\partial{\rm Re}\sigma^R(\omega)\over\partial\omega}
\right|_{\omega=\epsilon_k}.
\label{qp11}
\end{equation}

With respect to transport properties, the velocity renormalization
is the most important quasiparticle correction as it determines drift
of quasiparticles between collisions. The wave-function
renormalization as a function of the impurity potential is presented
in Fig.~\ref{fig:vq2}. There is a striking similarity of the
wave-function renormalization and the magnitude of virial corrections.

In the above discussion we have ignored selfconsistency. In Appendix~A
it is shown that for the weak scattering, ${1\over\tau}\to 0$,
the above formulas are identical to those resulting
from selfconsistent treatment. In Appendix~B we also derive a formula
that connects virial correction $1+\Delta_t/\tau$ with quasiparticle
renormalization $z$. This formula explains similar magnitudes of these
two corrections.

\section{Quasiparticle Boltzmann equation with collision delay}
Similarity of magnitudes of quasiparticle and virial corrections show
that both corrections have to be included in the transport equation
within the same accuracy. It is quite easy to guess such a transport
equation. The quasiparticle renormalization affects the drift between
collisions, therefore it enters the transport equation as a
renormalization of velocity (\ref{qp10}). The virial corrections enter
the scattering integrals like in (\ref{cl2}). The transport equation
that includes both corrections reads
\begin{eqnarray}
{\partial f\over\partial t}&+&z{k\over m}
{\partial f\over\partial r}-{\partial\phi\over\partial r}
{\partial f\over\partial k}=-{f\over\tau}
\nonumber\\
&+&{1\over\tau}{2\pi^2\over k^2}\int{dp\over(2\pi)^3}\delta(|p|-|k|)
f\left(p,r,t-\Delta_t\right).
\label{qp12}
\end{eqnarray}
Although this equation has the classical form (\ref{cl2}), its
components $z$, $\tau$, and $\Delta_t$ are given by quantum-mechanical
microscopic dynamics. One can also view (\ref{qp12}) as a phenomenologic
equation with momentum-dependent parameters $z$, $\Delta_t$ and $\tau$.

Beside the transport equation, one also needs relation of observables
to distribution function $f$. From the equation of continuity one
finds that the physical density includes only virial corrections,
\begin{equation}
n=\int{dk\over(2\pi)^3}\left(1+{\Delta_t\over\tau}\right)f,
\label{k9}
\end{equation}
while the density of particle current has only quasiparticle
corrections
\begin{eqnarray}
j&=&\int{dk\over(2\pi)^3}z {k\over m} f
\nonumber\\
&=&\int{dk\over(2\pi)^3}{\partial\varepsilon\over\partial k}f.
\label{k25}
\end{eqnarray}
>From the energy conservation one finds that the energy density includes
both corrections
\begin{equation}
E=\int{dk\over(2\pi)^3}\left(1+{\Delta_t\over\tau}\right)
(\varepsilon+\phi)f.
\label{k13}
\end{equation}
Both conservation laws are in Appendix~C.

>From the set of equations (\ref{qp12}-\ref{k13}) one can evaluate
properties of electron gas or liquid in a similar manner as one
uses the BE to this end. To demonstrate such an application, in the
next section we evaluate the dielectric function.

\section{Dielectric function}
The virial corrections influence a response of the system to
perturbations. The time non-locality of the scattering integral emerges
in non-stationary processes. The simplest but important process is
linear screening of external field described by dielectric function
$\kappa_r$.

The virial corrections enter the dielectric function in two ways, from
the transport equation (\ref{qp12}) and from functional (\ref{k9}). To
demonstrate both mechanisms, we evaluate $\kappa_r$ from its definition.

An electrostatic external potential
\begin{equation}
\phi_0(r,t)=\phi_0{\rm e}^{iqx-i\omega t}
\label{tc3a}
\end{equation}
creates a perturbation in the electron density
\begin{equation}
\tilde n(r,t)=\tilde n{\rm e}^{iqr-i\omega t}.
\label{tc5}
\end{equation}
The perturbation in density creates Coulomb potential
\begin{equation}
\tilde\phi(r,t)=\tilde\phi{\rm e}^{iqr-i\omega t}=
{e^2\over\kappa q^2}\tilde n{\rm e}^{iqr-i\omega t}
\label{tc5a}
\end{equation}
that adds to the external one so that the internal field reads
\begin{equation}
\phi=\phi_0+\tilde\phi=\phi_0+{e^2\over\kappa q^2}\tilde n.
\label{tc5b}
\end{equation}
Here, $\kappa$ is permitivity of the host crystal. From definition
\begin{equation}
\phi={\phi_0\over\kappa_r},
\label{ct5c}
\end{equation}
one finds the dielectric function to be
\begin{equation}
\kappa_r=1-{e^2\over\kappa q^2}{\tilde n\over\phi}.
\label{tc3}
\end{equation}

\subsection{Perturbation of quasiparticle distribution}
To evaluate perturbation $\tilde n$ of the physical density $n$, we
have to find the linear perturbation of the quasiparticle distribution,
\begin{equation}
\tilde f(k,r,t)=\tilde f(k){\rm e}^{iqr-i\omega t},
\label{tc6}
\end{equation}
caused by potential $\phi$. To this end we use linearized transport
equation (\ref{qp12})
\begin{eqnarray}
&&\left(-i\omega+iz{kq\over m}+{1\over\tau}\right)
\tilde f(k)-iq\phi{\partial f_0(k)\over\partial k}
\nonumber\\
&&={1\over\tau}{2\pi^2\over k^2}\int{dp\over(2\pi)^3}\delta(|p|-|k|)
{\rm e}^{i\omega\Delta_t}\tilde f(p).
\label{tc7a}
\end{eqnarray}
The momentum derivative of the equilibrium distribution
$f_0(k)=f_{FD}(\varepsilon_k)$ reads
\begin{equation}
{\partial f_0(k)\over\partial k}=z{k\over m}{\partial f_{FD}
(\varepsilon_k)\over\partial\varepsilon_k}.
\label{tc8}
\end{equation}

The perturbation $\tilde f$ depends only on the absolute value of
momentum $|k|$ and the angle between momentum $k$ and wave vector $q$.
We denote $s={kq\over |k||q|}$ and $s^\prime={pq\over |p||q|}$,
and integrate over the energy conserving $\delta$ function so that
the transport equation simplifies as
\begin{eqnarray}
\left(-i\omega+isz{|k||q|\over m}+{1\over\tau}\right)
&&\tilde f(|k|,s)-isz{|k||q|\over m}\phi{\partial f_{FD}
(\varepsilon_{|k|})\over\partial\varepsilon_{|k|}}
\nonumber\\
&&={1\over 2\tau}{\rm e}^{i\omega\Delta_t}\int_{-1}^1ds^\prime
\tilde f(|k|,s^\prime).
\label{tc9}
\end{eqnarray}
With abbreviations $z{|k|\over m}\equiv u$, $|q|\equiv q$, and skipping
argument $|k|$ in distributions, equation (\ref{tc7a}) reads
\begin{eqnarray}
\left(-i\omega+isqu+{1\over\tau}\right)\tilde f(s)&-&
isqu\phi{\partial f_{FD}\over\partial\varepsilon}
\nonumber\\
&=&{1\over 2\tau}{\rm e}^{i\omega\Delta_t}\int_{-1}^1ds^\prime
\tilde f(s^\prime).
\label{tc10}
\end{eqnarray}

The angular dependence of the distribution is easily found from
(\ref{tc10})
\begin{equation}
\tilde f(s)={isqu\phi{\partial f_{FD}\over\partial\varepsilon}+
{1\over\tau}{\rm e}^{i\omega\Delta_t}\tilde F
\over -i\omega+isqu+{1\over\tau}},
\label{tc12}
\end{equation}
where
\begin{equation}
\tilde F={1\over 2}\int_{-1}^1ds\tilde f(s),
\label{tc11}
\end{equation}
is an angle-averaged distribution.
Integrating over $s$, one finds from (\ref{tc12}) a condition for the
angle-averaged distribution
\begin{equation}
\tilde F=\left(1+(1-i\omega\tau)J\right)
\phi{\partial f_{FD}\over\partial\varepsilon}-
{\rm e}^{i\omega\Delta_t}J\tilde F,
\label{tc13}
\end{equation}
where
\begin{equation}
J={i\over 2qu\tau}{\rm ln}\left(
{\omega+{i\over\tau}-qu\over\omega+{i\over\tau}+qu}\right).
\label{tc14}
\end{equation}

Since the BE holds only for slowly varying fields, we can linearize
in $\Delta_t$, ${\rm e}^{i\omega\Delta_t}\approx 1+i\omega\Delta_t$.
The angle-averaged distribution from (\ref{tc13}) then results
\begin{equation}
\tilde F=\phi{\partial f_{FD}\over\partial\varepsilon}
{1+(1-i\omega\tau)J\over 1+(1+i\omega\Delta_t)J}.
\label{tc15}
\end{equation}
Now the perturbation of the quasiparticle distribution is fully
determined by (\ref{tc12}) and (\ref{tc15}).

\subsection{Perturbation of density}
Perturbation of the electron density is found from (\ref{k9})
\begin{eqnarray}
\tilde n&=&2\int{dk\over(2\pi)^3}\tilde f(k)
\nonumber\\
&=&{1\over\pi^2}
\int_0^\infty dk k^2 \tilde F(k)\left(1+{\Delta_t\over\tau}\right).
\label{tc16}
\end{eqnarray}
The factor of two stands for sum over spins.

For simplicity we assume the limit of low temperature
\begin{equation}
{\partial f_{FD}\over\partial\varepsilon}\to -\delta
(\varepsilon-E_F)=-{m\over zk}\delta(k-k_F),
\label{tc17}
\end{equation}
where one can easily integrate out the momentum
\begin{equation}
\tilde n=-\left.\phi{mk\over\pi^2z}\left(1+{\Delta_t\over\tau}\right)
{1+(1-i\omega\tau)J\over 1+(1+i\omega\Delta_t)J}\right|_{k=k_F}.
\label{tc18}
\end{equation}
Using (\ref{tc18}) in (\ref{tc3}) one directly obtains the dielectric
function.

\subsection{Long wave length limit}
Now we focus on long wave length limit, $q\to 0$. To evaluate
this limit from (\ref{tc18}) we first rearrange (\ref{tc14}) as
\begin{eqnarray}
J&=&-{1\over 1-i\omega\tau}
\sum_{x=\pm i{qu\tau\over 1-i\omega\tau}}{1\over x}\ln (1+x)
\nonumber\\
&\to &-{1\over 1-i\omega\tau}
\left[1-{1\over 3}\left({qu\tau\over 1-i\omega\tau}\right)^2\right].
\label{tc19}
\end{eqnarray}
In the long wave length limit, the dielectric function reads
\begin{equation}
\kappa_r=1+{{e^2mk_F\over\kappa\pi^2z}\left(1+{\Delta_t\over\tau}\right)
\over q^2{1+i\omega\Delta_t\over 1-i\omega\tau}-
3\omega\left(\omega+{i\over\tau}\right)
\left(1+{\Delta_t\over\tau}\right){m^2\over k_F^2z^2}},
\label{tc20}
\end{equation}
where $z$, $\tau$ and $\Delta_t$ are values at the Fermi level.

In the static case $\omega=0$, the dielectric function is of form
$\kappa_r=1+{q_s^2\over q^2}$. From (\ref{tc20}) one finds that
the Thomas-Fermi screening length $1/q_s$ is
\begin{equation}
q_s^2=
{e^2mk_F\over\kappa\pi^2}{1\over z}\left(1+{\Delta_t\over\tau}\right).
\label{tc21}
\end{equation}
Part ${e^2mk_F\over\kappa\pi^2}$ gives standard Thomas-Fermi screening,
factors ${1\over z}$ and $1+{\Delta_t\over\tau}$ provide quasiparticle
and virial corrections, respectively. As one can see from
Fig.~\ref{fig:vq2}, quasiparticle and virial corrections nearly equal,
therefore they mutually compensate in the Thomas-Fermi screening length
(\ref{tc21})
\begin{equation}
q_s^2\approx {e^2mk_F\over\kappa\pi^2}.
\label{tc21k}
\end{equation}

For homogeneous perturbations, $q=0$, the dielectric function is of
form $\kappa_r=1-{\omega_p^2\over\omega(\omega+i/\tau)}$. From
(\ref{tc20}), the plasma frequency $\omega_p$ results
\begin{equation}
\omega_p^2={e^2k_F^3\over3\kappa\pi^2 m}z.
\label{tc22}
\end{equation}
There is only the quasiparticle correction $z$.

Note that virial corrections to the Thomas-Fermi screening $q_s$ and
plasma frequency $\omega_p$ appear in rather paradoxical way. While
the static screening has virial corrections [due to $n[f]$, Eq.
(\ref{tc16})], the plasma frequency describing non-stationary
behavior has none. This is because in the homogeneous case, $q=0$,
the virial corrections from $n[f]$ and the scattering integral
mutually cancel due to particle conservation law.

\subsection{Virial correction to Fermi momentum}
The Thomas-Fermi screening length (\ref{tc21}) and the plasma frequency
(\ref{tc22}) are expressed in terms of the Fermi momentum. Additional
virial corrections to those quantities appear if one rewrite them
in terms of physical density $n$.

In general, the Fermi momentum is a parameter of the quasiparticle
distribution $f$, therefore it is always related to the free density.
For the parabolic band, the Fermi momentum results from (\ref{cl3}) as
\begin{equation}
k_F=\sqrt[3]{3\pi^2n_{\rm free}}.
\label{tc2}
\end{equation}
For sufficiently low density $n$, the ratio $\Delta_t/\tau$ changes
a little from zero to the Fermi momentum. In our case, $\Delta_t/\tau$
changes by 7\% for density $n=10^{16}$~cm$^{-3}$. This weak dependence
allows us to take approximation
\begin{equation}
{\Delta_t(k)\over\tau(k)}={\Delta_t(k_F)\over\tau(k_F)}.
\label{tc2a}
\end{equation}
>From (\ref{k9}) one finds that the free and physical densities
relate as
\begin{equation}
n=n_{\rm free}\left(1+{\Delta_t\over\tau}\right),
\label{tc2b}
\end{equation}
and the Fermi momentum reads
\begin{equation}
k_F=\sqrt[3]{{3\pi^2n\over 1+{\Delta_t\over\tau}}}.
\label{tc2c}
\end{equation}

In terms of the physical density, the Thomas-Fermi screening length
(\ref{tc21k}) regains corrections
\begin{equation}
q_s^2\approx {e^2m\over\kappa\pi^2}
\sqrt[3]{{3\pi^2n\over 1+{\Delta_t\over\tau}}}.
\label{tc21a}
\end{equation}
In opposite, the plasma frequency in terms of physical density
regains its free-particle value
\begin{eqnarray}
\omega_p^2&=&{e^2n\over\kappa m}{z\over 1+{\Delta_t\over\tau}}
\nonumber\\
&\approx &{e^2n\over\kappa m}.
\label{tc22a}
\end{eqnarray}

\subsection{dc conductivity}
Compensation of virial and quasiparticle corrections also appears for
dc conductivity $\sigma_{\rm dc}$. Although this compensation is a
direct consequence of the dielectric function, we discuss it in detail
for its experimental importance.

The conductivity relates to dielectric function $\kappa_r$ as
\begin{equation}
\sigma_{\rm dc}=\lim_{q,\omega\to 0}-{i\omega\kappa}(\kappa_r-1).
\label{dc1}
\end{equation}
This known relation can be recovered from the equation of continuity
(\ref{k27}) that yields $i\omega\tilde n-iqj=0$, where $j$ is flow
of particles. The electric field $F$ results from the
electrostatic potential as $eF=iq\phi$. The conductivity then reads
\begin{equation}
\sigma_{\rm dc}=-{ej\over F}=i{e^2\omega\over q^2}{\tilde n\over\phi}.
\label{dc2}
\end{equation}
Comparing (\ref{dc2}) with (\ref{tc3}) one recovers (\ref{dc1}).

Sending $q\to 0$ and $\omega\to 0$ one finds standard relaxation
time formula with the quasiparticle correction
\begin{equation}
\sigma_{\rm dc}={e^2k_F^3\tau\over 3\pi^2m}z.
\label{dc3}
\end{equation}
In terms of the physical density,
\begin{eqnarray}
\sigma_{\rm dc}&=&{e^2n\tau\over m}{z\over 1+{\Delta_t\over\tau}}
\nonumber\\
&\approx &{e^2n\tau\over m},
\label{dc4}
\end{eqnarray}
virial and quasiparticle corrections mutually compensate.

\section{Summary}
We have shown that for scattering by resonant levels of neutral
impurities the virial and quasiparticle corrections are of the
same magnitude. We have proposed an intuitive modification of the BE
that includes both corrections. Proposed modification of the BE has
quasiparticle corrections in the drift term (as in the Landau theory)
and virial corrections in the scattering integral (as in the classical
theory of dense gases). The modified BE can be solved as simply as the
standard BE.

An interplay of virial and quasiparticle corrections has been discussed
on the dielectric function. Various compensations of virial and
quasiparticle corrections has been demonstrated on the static screening,
plasma frequency and dc conductivity. Careful measurements of the
dielectric function in III-V semiconductor with resonant levels can
reveal this interplay. Sensitivity of resonant levels to a hydrostatic
pressure makes possible to control magnitude of virial and quasiparticle
corrections in a single sample.

The way we have introduced the transport equation (\ref{qp12}) does
not guarantee its validity. There are two fundamental questions one
has to ask:
(i) Has the time non-locality of the quantum-mechanical scattering
integral really the same form as the classical collision delay?
(ii) Are quasiparticle and virial corrections included in consistent
manner?
No intuitive argument can give satisfactory answers to these questions.
To prove yes-answers for both questions one has to recover the
transport equation from quantum statistics. Such a microscopic theory is
in the second paper of this sequence.

\acknowledgments
This work was supported from the Grant Agency of the Czech Republic
under contract Nr. 202960098, the BMBF (Germany) under contract Nr.
06R0745(0), and the EC Human Capital and Mobility Programme.
\appendix
\section{Self-consistent ATA}
Here we show that for weak scattering, ${1\over\tau}\to 0$,
non-self-consistent formulas from Sec.~VB result also from the
self-consistent generalization of the self-energy.

The effect of impurities on the band structure affects propagation of
the electron in the region between impurities. This change should be
included in the propagator that enters the T-matrix. Instead of
the unperturbed propagator $G^R_0$, the T-matrix in the self-energy
(\ref{qp5}) should be constructed from the full propagator $G^R$
\begin{equation}
t^R_{\rm self}=v+V\langle 0|G^R|0\rangle t^R_{\rm self},
\label{qp8a}
\end{equation}
where $G^R$ is given by the Dyson equation
\begin{equation}
G^R=G^R_0+G^R_0\Sigma_{\rm self}^RG^R,
\label{qp8b}
\end{equation}
and
\begin{equation}
\sigma^R_{\rm self}=ct^R_{\rm self}.
\label{qp8s}
\end{equation}
This approximation is called the self-consistent ATA.\cite{VKE68}

For the KS impurity, where the self-energy has no momentum dependence,
the self-consistency is simply achieved by a complex shift of the
energy argument
\begin{equation}
G^R(\omega)=G^R_0(\omega-\sigma^R_{\rm self}).
\label{qp8c}
\end{equation}
>From (\ref{qp8c}) one finds that
$t^R_{\rm self}(\omega)=t^R(\omega-\sigma^R_{\rm self})$, therefore
\begin{equation}
\sigma^R_{\rm self}(\omega)=\sigma^R(\omega-\sigma^R_{\rm self}).
\label{qp8c1}
\end{equation}

\subsection{Energy}
Within the self-consistent treatment the quasiparticle energy is
defined as\cite{SL95}
\begin{equation}
\varepsilon_k=\epsilon_k+{\rm Re}\sigma_{\rm self}^R(\varepsilon_k).
\label{qp8c2}
\end{equation}
Using (\ref{qp8c1}), one can rewrite (\ref{qp8c2}) as
\begin{equation}
\varepsilon_k=\epsilon_k+
{\rm Re}\sigma^R(\varepsilon_k-{\rm Re}\sigma^R_{\rm self}).
\label{qp8c3}
\end{equation}
In argument of the self-energy we use definition (\ref{qp8c2}) to
recover
$$\varepsilon_k=\epsilon_k+{\rm Re}\sigma^R(\epsilon_k).
\eqno(\ref{qp7})$$

\subsection{Wave-function renormalization}
Within the self-consistent treatment, the wave-function renormalization
results as\cite{SL95}
\begin{equation}
z={1\over 1-\left.{\partial{\rm Re}\sigma^R_{\rm self}\over
\partial\omega}\right|_{\omega=\varepsilon_k}}.
\label{qp8c4}
\end{equation}
>From (\ref{qp8c1}) follows
\begin{equation}
\left.{\partial{\rm Re}\sigma^R_{\rm self}\over
\partial\omega}\right|_{\omega=\varepsilon_k}=\left.{
\partial{\rm Re}\sigma^R\over\partial\omega}
\right|_{\omega=\epsilon_k}
\left(1-\left.{\partial{\rm Re}\sigma^R_{\rm self}\over
\partial\omega}\right|_{\omega=\varepsilon_k}\right),
\label{qp8c5}
\end{equation}
which can be rewritten as
\begin{equation}
{1\over 1-\left.{\partial{\rm Re}\sigma^R_{\rm self}\over
\partial\omega}\right|_{\omega=\varepsilon_k}}=
1+\left.{\partial{\rm Re}\sigma^R\over\partial\omega}
\right|_{\omega=\epsilon_k}.
\label{qp8c6}
\end{equation}
The wave-function renormalization (\ref{qp8c4}) is thus identical to
(\ref{qp11}).

\subsection{Lifetime}
Within the self-consistent treatment, the inverse lifetime results as
\cite{SL95}
\begin{equation}
{1\over\tau}=z{\rm Im}\sigma^R_{\rm self}(\varepsilon_k).
\label{qp8c7}
\end{equation}
>From (\ref{qp8c1}) and (\ref{qp8c2}) one finds
\begin{eqnarray}
\sigma^R_{\rm self}(\varepsilon_k)&=&
\sigma^R(\varepsilon_k-{\rm Re}\sigma^R_{\rm self}-i{\rm Im}
\sigma^R_{\rm self})
\nonumber\\
&=&\sigma^R(\epsilon_k-i{\rm Im}\sigma^R_{\rm self})
\nonumber\\
&=&\sigma^R(\epsilon_k)-i{\rm Im}\sigma^R_{\rm self}(\varepsilon_k)
\left.{\partial\sigma^R\over\partial\omega}\right|_{\omega=\epsilon_k}.
\label{qp8c8}
\end{eqnarray}
In the last line we have used linear approximation in the imaginary
part of the argument which holds for weak scattering,
${1\over\tau}\to 0$.

Imaginary part of (\ref{qp8c8})
\begin{equation}
{\rm Im}\sigma^R_{\rm self}(\varepsilon_k)={\rm Im}\sigma^R(\epsilon_k)-
{\rm Im}\sigma^R_{\rm self}(\varepsilon_k)
\left.{\partial{\rm Re}\sigma^R\over\partial\omega}
\right|_{\omega=\epsilon_k}
\label{qp8c9}
\end{equation}
can be rearranged as
\begin{equation}
z{\rm Im}\sigma^R_{\rm self}(\varepsilon_k)=
{\rm Im}\sigma^R(\epsilon_k).
\label{qp8c10}
\end{equation}
Formula (\ref{qp8c7}) is thus identical to (\ref{qp5b}).

\section{Quasiparticle versus virial corrections}
Although virial and quasiparticle corrections describe different
features of the quasiparticle transport, both of them are linked to
energy derivatives of the T-matrix. From this link follows similarity
of their magnitudes.

One can rearrange formula (\ref{ct5}) in the way that reveals the
relation of virial correction $1+{\Delta_t\over\tau}$ to wave-function
renormalization $z$. Writing (\ref{ct5}) as $\Delta_t={\rm Im}
{1\over t^R}{\partial t^R\over\partial\omega}={1\over 2i}\left(
{1\over t^R}{\partial t^R\over\partial\omega}-
{1\over t^A}{\partial t^A\over\partial\omega}\right)$ and the inverse
lifetime (\ref{ct7}) as ${1\over\tau}=2c{\rm Im}t^R=ic(t^R-t^A)$,
[$t^A$ is complex conjugate to $t^R$], one obtains that
\begin{eqnarray}
{\Delta_t\over\tau}
&=&ic(t^R-t^A){1\over 2i}\left(
{1\over t^R}{\partial t^R\over\partial\omega}-
{1\over t^A}{\partial t^A\over\partial\omega}\right)
\nonumber\\
&=&{\partial\over\partial\omega}{c\over 2}(t^R+t^A)-{c\over 2}
\left({t^A\over t^R}{\partial t^R\over\partial\omega}-
{t^R\over t^A}{\partial t^A\over\partial\omega}\right)
\nonumber\\
&=&{\partial{\rm Re}\sigma^R\over\partial\omega}-{c\over 2}
\left({t^A\over t^R}{\partial t^R\over\partial\omega}-
{t^R\over t^A}{\partial t^A\over\partial\omega}\right),
\label{qvv1}
\end{eqnarray}
where we have used that ${\rm Re}\sigma^R={c\over 2}(t^R+t^A)$. From
(\ref{ts7}) one finds
\begin{equation}
{\partial t^R\over\partial\omega}=t_R^2
{\partial\over\partial\omega}\langle 0|G^R_0|0\rangle ,
\label{qvv2}
\end{equation}
which substituted into (\ref{qvv1}) provides
\begin{equation}
{\Delta_t\over\tau}={\partial{\rm Re}\sigma^R\over\partial\omega}-
c|t^R|^2{\rm Re}{\partial\over\partial\omega}\langle 0|G^R_0|0\rangle .
\label{qvv3}
\end{equation}

Formula (\ref{qvv3}) makes connection between virial and quasiparticle
corrections. One can see that at least two limiting regimes can be
distinguished according to relative values of the first and the
second terms in (\ref{qvv3}).

For weak potentials when the self-energy can be treated in the Born
approximation $t^R\approx v$, i.e. $\sigma^R=cv^2\langle 0|G^R_0|0
\rangle$, the virial corrections vanish because the first and the
second terms mutually cancel. In contrast, the quasiparticle
corrections remain. Since most of quasiclassical
transport equations have been derived within the Born approximation
(or single-loop approximation for particle-particle interaction), it
is quite natural that they do not include virial corrections.

The scattering by resonant levels is far from the Born approximation,
for model and parameters we consider here $|t^R|\sim 100\times|v|$.
In this case, the second term of (\ref{qvv3}) is of the order of
$10^{-3}$ while the first one is of the order of $10^{-1}$. Accordingly,
the second term can be neglected, i.e., virial and quasiparticle
corrections are of the same magnitude.

\section{Conservation laws}
\subsection{Equation of continuity}
In inhomogeneous and non-stationary system, there are currents $j$ due
to which local density of electrons $n$ changes. Here we prove that
density (\ref{k9}) and current (\ref{k25}) are consistent with the BE
(\ref{qp12}) obeying the equation of continuity.

Under integration over momentum, term ${\partial f\over\partial t}$
turns into ${\partial n_{\rm free}\over\partial t}$, and the scattering
integrals turn into $-{\partial n_{\rm corr}\over\partial t}$, see
Sec.~II. Equation (\ref{qp12}) then yields
\begin{equation}
{\partial n\over\partial t}+\int{dk\over(2\pi)^3}
\left(z{k\over m}{\partial f\over\partial r}-
{\partial\phi\over\partial r}{\partial f\over\partial k}\right)=0 .
\label{k26}
\end{equation}
The second term in the brackets vanishes what follows from its
integration by parts,
\begin{equation}
{\partial n\over\partial t}+{\partial\over\partial r}
\int{dk\over(2\pi)^3}z{k\over m} f=0.
\label{k27}
\end{equation}
The second term is the divergency of the current $j$ given by
(\ref{k25}). Equation (\ref{k27}) is thus the equation of continuity.

\subsection{Conservation of energy}
Here we prove that for homogeneous system the energy of electrons
(\ref{k13}) changes with the field $\phi$ in a consistent way, i.e.,
${\partial E\over\partial t}=n{\partial\phi\over\partial t}$. First
we take time derivative of equation (\ref{k13}),
\begin{eqnarray}
{\partial E\over\partial t}&=&\int{dk\over(2\pi)^3}
{\partial\phi\over\partial t}f\left(1+{\Delta_t\over\tau}\right)
\nonumber\\
&+&\int{dk\over(2\pi)^3}(\varepsilon+\phi){\partial f\over\partial t}
\left(1+{\Delta_t\over\tau}\right).
\label{k15}
\end{eqnarray}

The second term of (\ref{k15})
vanishes. To show this, we multiply the BE (\ref{qp12})
with the quasiparticle energy $\varepsilon+\phi$ and integrate over
momentum
\begin{equation}
\int{dk\over(2\pi)^3}(\varepsilon+\phi){\partial f\over\partial t}=-
\int{dk\over(2\pi)^3}(\varepsilon+\phi){\partial f\over\partial t}
{\Delta_t\over\tau}.
\label{k14}
\end{equation}
We have used that contributions of non-gradient terms of the
scattering integrals mutually cancel because of the energy-conserving
$\delta$ function.

In the first term of (\ref{k15}) we take out the field, the rest of the
integral is just density (\ref{k9}). We have thus proved that the
total energy changes in thermodynamically consistent way
\begin{equation}
{\partial E\over\partial t}=n{\partial\phi\over\partial t}.
\label{k21}
\end{equation}

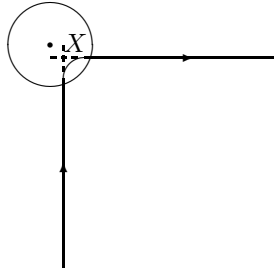
\begin{figure}
\begin{picture}(200,130)(-100,-100)
\put(0,0){\circle*{2}}
\put(0,0){\circle{30}}
\put(5,-14.33){\line(0,-1){30}}
\put(5,-84.33){\vector(0,1){40}}
\put(14.33,-5){\vector(1,0){40}}
\put(54.33,-5){\line(1,0){30}}
\put(14.33,-14.33){\oval(18.66,18.66)[tl]}
\multiput(5,-10.33)(0,4){3}{\line(0,1){2}}
\multiput(10.33,-5)(-4,0){3}{\line(-1,0){2}}
\put(5,-3){{\normalsize $X$}}
\end{picture}
\caption{Scheme of classical collision. The circle represents limits
of an impurity potential, the full line is an electron trajectory, the
dashed lines are extrapolations of incoming and outgoing trajectories.}
\label{fig:cl1}
\end{figure}

\begin{figure}
\begin{picture}(200,130)(-100,-100)
\put(0,0){\circle*{2}}
\put(0,0){\circle{30}}
\put(10.52,-10.52){\line(0,-1){30}}
\put(10.52,-80.52){\vector(0,1){40}}
\put(10.52,-10.52){\vector(1,0){40}}
\put(50.52,-10.52){\line(1,0){30}}
\multiput(0,0)(0,-4){21}{\line(0,-1){2}}
\multiput(0,0)(4,0){21}{\line(1,0){2}}
\put(5,-60){\vector(-1,0){5}}
\put(5,-60){\vector(1,0){5.51}}
\put(4,-58){$b$}
\put(0,-75){\circle*{3}}
\put(10.52,-75){\circle*{3}}
\put(-10,-78){$\bar{\rm A}$}
\put(13,-78){A}
\put(78,0){\circle*{3}}
\put(78,-10.52){\circle*{3}}
\put(75,3){$\bar{\rm B}$}
\put(75,-20){B}
\put(0,0){\vector(-2,1){15}}
\put(-5,5){$R$}
\end{picture}
\caption{Collision with a hard sphere. The full line is a real electron
trajectory, the dashed lines are effective trajectories used within the
BE.}
\label{fig:cl2}
\end{figure}
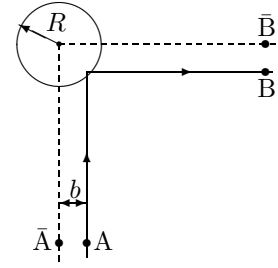

\begin{figure}
\epsfxsize=9cm
\epsffile{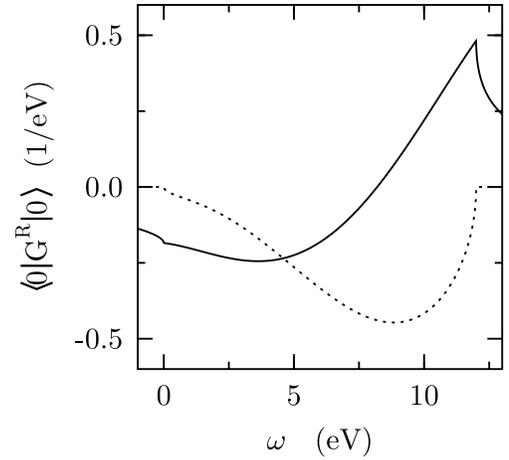}
\caption{Local Green's function. The imaginary part of local Green
function (dotted line) has a low density of state at the band
edge $\omega\sim 0$ corresponding to low effective mass $m=0.2$,
and high shoulder at higher energies simulating for satellite
minima. The real part (full line) has finite value
$\sim -0.185$~1/eV and is
nearly flat at the vicinity of the band edge.}
\label{fig:ts1}
\end{figure}

\begin{figure}
\epsfxsize=9cm
\epsffile{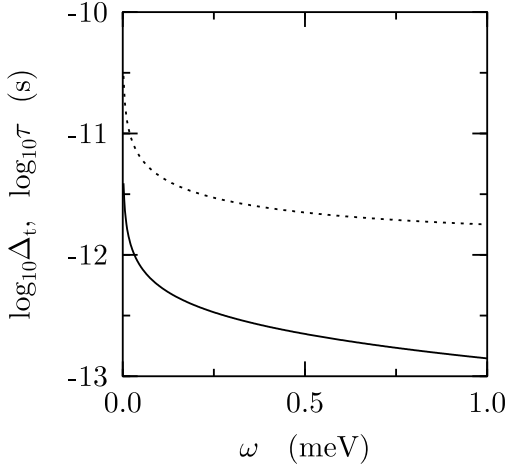}
\caption{Collision delay and lifetime as function of energy.
Except for a shift by an order of magnitude, the collision delay
(full line) has nearly the same energy dependence as the lifetime
(dashed line).}
\label{fig:ct1}
\end{figure}

\begin{figure}
\epsfxsize=9cm
\epsffile{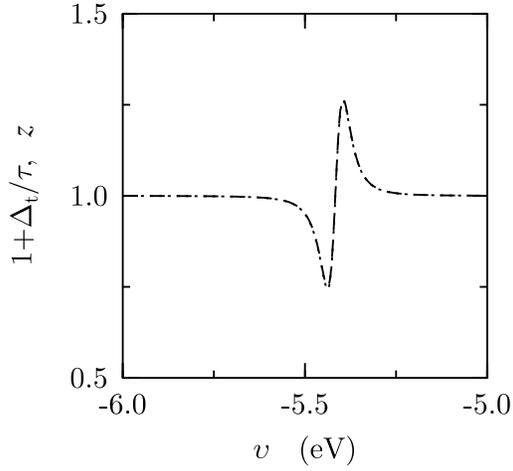}
\caption{Virial and quasiparticle corrections for electron energy
100~meV above band edge and concentration $c=10^{-6}$. Virial
correction (dashed line) is greater than 1 for resonant levels,
$v>-5.4$~eV, what corresponds to positive collision delay.
Quasiparticle renormalization $z$ (dotted line) nearly equals
the virial correction. In fact they differ less than by 0.8\%.}
\label{fig:vq2}
\end{figure}

\begin{figure}
\epsfxsize=9cm
\epsffile{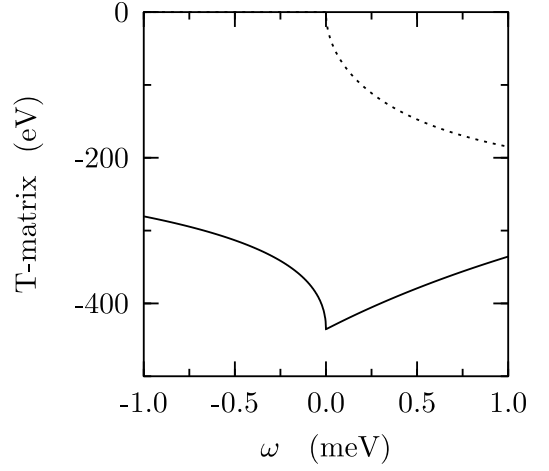}
\caption{T-matrix near band edge. The real part ${\rm Re}t^R$ is
in full line, the imaginary part ${\rm Im}t^R$ is in dashed line.}
\label{fig:ts2}
\end{figure}

\begin{figure}
\epsfxsize=9cm
\epsffile{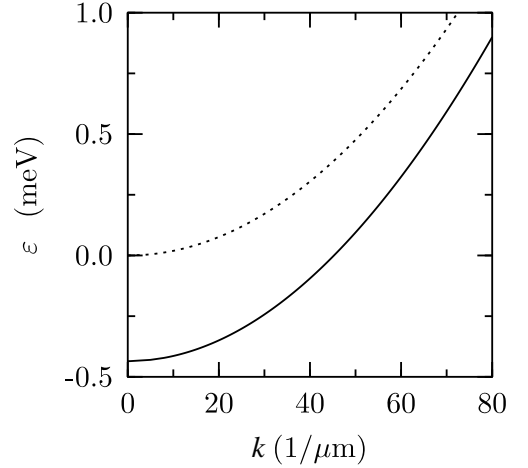}
\caption{Quasiparticle energy (full line) for resonant levels,
$v=-5.35$~eV, of concentration $c=10^{-6}$. Bare kinetic energy
$\epsilon_k$ (dotted line) serves as an eye guide.}
\label{fig:qp1}
\end{figure}
\end{document}